\documentclass{aa}  
\usepackage{booktabs}
\usepackage{lipsum}
\usepackage{amsmath} 
\usepackage[utf8]{inputenc}
\usepackage{graphicx}
\usepackage{txfonts}
\usepackage{natbib}
\setcitestyle{authoryear,open={},close={}}

\DeclareUnicodeCharacter{2212}{-}

\begin{document}

   \title{Exploring the IR-radio correlation in massive galaxy clusters at the end of cosmic noon}

   \author{N. Samanso
          \inst{1},
          J.B. Nantais\inst{1},
          S. Alberts\inst{2},
          U. Rescigno\inst{3},
          W. Rujopakarn\inst{4,5},
          G. R. Zeimann\inst{6},
          \and
          J. Wagg\inst{7}
          }

   \institute{Instituto de Astrofisica, Departamento de Fisica, Facultad de Ciencias Exactas, Universidad Andres Bello, Fernandez Concha 700, Las Condes, Santiago RM, Chile\\
              \email{n.samanso@uandresbello.edu}
         \and
             Steward Observatory, University of Arizona, 933 N. Cherry Ave., Tucson, AZ 85721, USA 
        \and
            Instituto de Astronomía y Ciencias Planetarias de Atacama, Universidad de Atacama, Copayapu 485, Copiapó, Chile
        \and
            National Astronomical Research Institute of Thailand, Don Kaeo, Mae Rim, Chiang Mai 50180, Thailand
        \and
            Department of Physics, Faculty of Science, Chulalongkorn University, 254 Phayathai Road, Pathumwan, Bangkok 10330, Thailand
        \and
            Hobby-Eberly Telescope, University of Texas at Austin, 2515 Speedway Boulevard, Austin, TX, 78712, USA
        \and
            Observatoire de la Côte d'Azur, CNRS, Laboratoire Lagrange, Bd de l'Observatoire, CS 34229, 06304, Nice Cedex 4, France
            }
        
   \date{Received  November 11, 2024; April 3, 2025}

  \abstract
  {Galaxies in overdense environments, such as clusters, present predominantly  quenched populations up to z $\sim$ 1. This suggests that environmental mechanisms suppress the star formation in these galaxies for the majority of cosmic history. At low redshifts, galaxies in rich group and cluster environments frequently show a radio emission excess in the IR-radio correlation, but this has yet to be confirmed at high redshifts, when environmental effects begin to strongly affect the evolution of cluster galaxies.}
   {We investigate the effect of the environment on the infrared and radio emission of cluster galaxies during the transition epoch at $1 < z < 2$ when they first start to quench consistently in the majority of galaxy clusters. }
   {We considered a sample of 129 cluster member galaxies from 11 massive clusters at a confirmed redshift of 1.0-1.8 from the IRAC Shallow Cluster Survey (ISCS), the IRAC Distant Cluster Survey (IDCS), and new 3 GHz images from the Karl G. Jansky Very Large Array (VLA). We calculated the IR-radio correlation slope parameter, q, in order to identify differences in the ratios of IR to radio of cluster galaxies and field galaxy comparison samples at different redshifts. Active galactic nuclei (AGNs) were identified and analyzed to search for any effect on the IR-radio correlation. The correlation parameter values were also compared by the Kolmogorov–Smirnov test with field galaxies.  }
   {Our comparison of the IR-radio correlation in cluster galaxies to the control sample of field galaxies reveals a marginally to modestly significant difference in the correlation slope parameter at the $ \sim2  \ \sigma-3\ \sigma$ level. A split of the clusters into low-redshift ($1<z<1.37$) and high-redshift ($1.37<z<1.8$) bins indicates a more significant difference in the correlation parameter in the lower-redshift cluster subsample, where widespread quenching begins. We find no difference in the IR-radio correlation between the high-z cluster and field samples, which is consistent with environmental effects on galaxy properties. These are less consistently observed at $z \gtrsim 1.4$. Our results suggest no evidence of any difference in the IR-radio correlation between galaxies that lie closer to the cluster center (projected radius R $< 1$ Mpc) and galaxies in the cluster outskirts (R $> 1$ Mpc). We find no difference in the IR-radio correlation between galaxies that host AGNs and non-active star-forming galaxies either. This suggests that our AGNs are overwhelmingly radio quiet and therefore do not affect the results we described above. We conclude that further investigations based on larger datasets are needed to constrain the impact of the cluster environment on the IR-radio correlation better. }
  {}


    \titlerunning{Exploring the IR-Radio Correlation in Massive Clusters at the End of Cosmic Noon}
    \authorrunning{N. Samanso, J.B. Nantais, S. Alberts, U. Rescigno et al.}
    
    \maketitle

\section{Introduction}

    A correlation between infrared (IR) and radio emission was first discovered by \citet{1971A&A....15..110V}. The correlation was later confirmed in other studies (e.g., \citet{1972BAAS....4Q.223L}, \citet{1973A&A....29..263V}, and \citet{1991ApJ...376...95C}). The IR-radio correlation empirically links the IR and radio emission in star-forming galaxies (SFGs) to processes related to the formation and destruction of short-lived massive stars. The ultraviolet and optical emission from young massive stars heats H{\sc II} regions and diffuse dust in the interstellar medium (ISM) of galaxies, which causes the re-emission of stellar radiation at IR wavelengths as determined by the dust temperature. Radio emission is generated by cosmic rays and electrons that are accelerated by the magnetic fields of core-collapse supernovae (SNe) remnants. This process is known as synchrotron radiation. 
    
    The correlation parameter q describes the slope of the IR-radio correlation in terms of the logarithm of the ratio of the IR-to-radio flux. It was originally defined by \citet{1985ApJ...298L...7H}. Later, \citet{2001ApJ...554..803Y} showed that q is constant over five magnitudes of galaxy luminosity in the low-redshift universe. Further studies, for example, \citet{2010A&A...518L..31I}, \citet{2010MNRAS.409...92J}, and \citet{2010ApJ...714L.190S}, found evidence that this relation is constant up to high redshift. The IR-radio correlation has several important applications. It can be used to identify radio-loud AGNs by measuring the galaxy offset from the correlation (\citep{2005ApJ...634..169D}, \citep{2013A&A...549A..59D}, \citep{2015MNRAS.453.1079B}, \citep{2017A&A...602A...3D}), as an assumption to determine the dust temperature and the redshift of submillimeter galaxies at high redshift (\citep{2005ApJ...622..772C}), and as a calibrator of radio-continuum emission that traces the star formation rate (SFR) without any bias from dust (\citep{1992ARA&A..30..575C}, \citep{2003ApJ...586..794B}, \citep{2011ApJ...737...67M}, \citep{2012ApJ...761...97M}).
    
     Various studies have recently shown that the value of the IR-radio correlation parameter weakly depends on redshift (e.g., \citet{2021MNRAS.504..118M}, \citet{2022MNRAS.511.1408G}). The simulation-based study of the history of the impacts of galaxy-galaxy interactions on the correlation parameter by \citet{2019MNRAS.489.4557P} also found evidence of a redshift evolution of the correlation parameter for disk and irregular SFGs. This supports the results found through the observations.
     
     Specific classes of galaxies such as distant starbursts and submillimeter galaxies (\citep{2006ApJ...650..592K}; \citep{2008ApJ...683..659S}; \citep{2010A&A...518L.154S}; \citep{2015A&A...576A.127S}), interacting galaxies (e.g., \citep{2013ApJ...777...58M}; \citep{2015MNRAS.453..638D}), and members of rich clusters (e.g., \citep{1995AJ....109.1582A}; \citep{2004ApJ...600..695R}; \citep{2009ApJ...706..482M}) may deviate from this relation. Several studies in the local Universe (e.g.,\citep{1995AJ....109.1582A}; \citep{2004ApJ...600..695R}; \citep{2009ApJ...706..482M}) showed that the value of the IR-radio correlation parameter in richer environments, such as galaxy clusters and massive groups, is lower than in the field. An excess in radio emission is thought to be the source of this deviation, and not a shortage in the IR emission. This radio excess was found to increase with increasing redshift (\citep{2004ApJ...600..695R}, \citep{2009ApJ...706..482M}). However, the study of the IR–radio correlation in a massive intermediate-redshift galaxy cluster by \citet{2015MNRAS.447..168R} showed no evidence of a change in the correlation value with either redshift or environmental effects. Nevertheless, it is noteworthy that this study was based on a single galaxy cluster. 
     
     The origin of radio excess is presumably associated with processes that drive the galaxy evolution in clusters, for example, galaxy interactions, mergers, or ram pressure stripping (e.g., \citep{2006PASP..118..517B}). Evidence from subkiloparsec studies of local cluster galaxies suggested compression from ram pressure on a galaxy scale (\citep{2009ApJ...706..482M}; \citep{2010A&A...512A..36V}). The compression results in asymmetries in the radio emission. However, \citep{2013A&A...553A.116V} showed that ram pressure stripping was not strong enough to explain the asymmetric radio continuum emission in several galaxies. This suggests effects not only from ram pressure, but also from galaxy interactions or recent minor mergers. These pieces of evidence indicate that the IR-radio correlation is an important tool for identifying and studying environment-driven galaxy evolution. 
     
     Understanding galaxy evolution requires a detailed analysis of the galaxy properties and their correlation with the environment, not only in the local Universe, but also throughout cosmic time. It is well known that the relation of the SFR in the local Universe is anticorrelated with the galaxy density (\citep{2007A&A...468...33E}; \citep{2008MNRAS.383.1058C}; \citep{2011A&A...532A.145P}). Passively evolving early-type galaxies (ETGs) tend to be found at the cores of massive galaxy clusters, while actively evolving SFGs are preferentially found in low- to intermediate-density environments, such as low-mass groups, or in the field. However, the correlation of star formation to density is only tight up to a redshift (z) of $\sim 1$, after which it becomes weak or absent in late cosmic noon (z $\sim 1 - 2$; \citealt{2013ApJS..206....3S}). Studies of particular clusters at $z > 1.4$ and $z > 1.6$ (by \citep{2010ApJ...719L.126T} and \citep{2010ApJ...718..133H}, respectively) showed an increase in the SFG fractions and specific SFRs (sSFR; where $sSFR=SFR/M_{\ast})$ in cluster cores. These results are similar to the evolution of the star formation rate with redshift reported for field galaxies (e.g., \citep{2007ApJ...670..156D}). 
     
     Based on data from the IRAC Shallow Cluster Survey (ISCS; \citep{2010ApJ...720..284M}), evidence was reported that galaxies in clusters are generally still assembling their stellar mass at redshift $z $>$ 1.3$, but evolve passively at lower redshifts ($z < 1.3$). \citet{2013ApJ...768....1M} and \citet{2016ApJ...825...72A} also revealed an increase in the AGN fractions in clusters with increasing redshift based on the same cluster sample. \citet{2016ApJ...825...72A} showed that the average SFR and sSFR in the ISCS clusters were suppressed compared to field galaxies at $z < 1.4$, but then became similar to the field at $z > 1.4$. These studies used stellar-mass selected cluster samples that were not biased toward evolved cluster populations to show that clusters at $z \sim 1-2$ are in a transition phase in which widespread environmental quenching was recently established.  
     
     We focus on the infrared and radio properties of galaxies in a sample of 11 massive $(> 10^{14} M_{\odot})$ spectroscopically confirmed galaxy clusters at $z = 1-1.75$. Ten of these clusters are from the ISCS at $z = 1-1.5$, and the remaining cluster is from the IDCS at $z = 1.75$. These 11 clusters were observed with deep confusion-limited \textit{Herschel}/PACS imaging. Considering deep far-IR observation at $z > 1$, \citet{2016ApJ...825...72A} provided the largest sample of massive galaxy clusters whose star formation related properties can be observed and analyzed in a manner analogous to field galaxy samples at cosmic noon ($z\sim 1-2$). Their far-IR observations were performed in order to effectively estimate the star formation in the typical heavily dust-obscured mode that dominates at cosmic noon. They also combined \textit{Herschel}/PAC data with multiwavelength data to quantify the UV to far-IR spectral energy distributions (SEDs) of galaxy clusters, which allowed them to calculate SFRs and sSFRs and to identify AGNs. It is worth mentioning that the breakdown in the SFR-density relation is observed in our sample of 11 clusters with deep IR data. This makes the results more robust than studies of individual clusters compared to field samples and galaxy clusters from other studies with inhomogeneous methods of the data analysis and collection. 

    Because studies of the IR-radio correlation performed in relatively homogeneously analyzed samples of multiple galaxy clusters instead of single clusters are rare, especially at high redshift, we provide results to clarify our understanding of the following scientific questions: We determine 1) whether differences in the environment affect the correlation values, 2) the mechanisms that might explain the tightness in the correlation for these clusters, whose galaxies are just beginning to quench, and 3) whether cluster galaxies are especially inclined to deviate significantly from the correlation. To do this, we investigated the similarities and differences in the IR-radio correlation as a function of redshift between clusters and the field and within clusters at different cluster-centric distances using 11 massive far-IR selected clusters at the end of cosmic noon. This time period is crucial in the galaxy evolution. 

    In Sect. \ref{data} we describe our sample and the data reduction. We identify the AGNs in Sect. \ref{AGN_iden}. In Sect. \ref{IR_correlation} we present our analysis of the IR-radio correlation as a function of redshift and compare it to noncluster-control samples. In Sect. \ref{presence_AGN} we present the result from our study by considering the influence (or lack thereof) of AGNs. In Sect. \ref{discussion} we discuss our results with respect to previous studies. We present our discussion and conclusions in Sect. \ref{discussion}. Throughout this work, we assume $ H_0 = 70 kms^{-1}Mpc^{-1}$, $\Omega_M = 0.3$, and $\Omega_{\Lambda} = 0.7$.


\section{Data}
\label{data}

    \subsection{IRAC Shallow and Distant Cluster Surveys}

    In the ISCS program, more than 300 galaxy cluster candidates at $z \sim 0.1-2$ were identified using IR-selected galaxy catalogs that covered 8.5 square degrees on the sky in the Boötes field. At $z > 1$, 20 galaxy clusters and rich groups from more than 100 cluster candidates were confirmed spectroscopically (\citep{2005sptz.prop20593S}; \citep{2006ApJ...651..791B, 2011ApJ...732...33B, 2013ApJ...779..138B}; \citep{2011A&A...533A.119E}; \citep{2008ApJ...684..905E}). A deeper follow-up survey, the IDCS, spectroscopically confirmed 2 clusters at z $\sim$ 1.8. Systems that are better classified as rich groups or protoclusters are expected to show less significant differences from the field at these redshifts. We focused on 11 spectroscopically confirmed massive galaxy clusters with halo masses ($M_{200}$) greater than $10^{14}$ \(M_\odot\) and deep \textit{Herschel}/PACS imaging (\citep{2016ApJ...825...72A}). The halo masses were determined from X-ray observations (\citep{2016ApJ...817..122B}) and weak lensing (\citep{2011ApJ...737...59J}). The clusters, along with their redshifts, positions, numbers of detected galaxies, and significance of detection, are listed in Table \ref{targets_list}. 

    Follow-up observations were performed on detected cluster galaxies to obtain spectroscopic redshifts using multi-object Keck Low Resolution Imaging Spectrograph (LRIS) optical spectroscopy and Wide Field Camera 3 (WFC3) slitless near-IR grism spectroscopy from the \textit{Hubble} Space Telescope (HST; see \citep{2013ApJ...779..138B}, \citep{2013ApJ...779..137Z} for a more detailed description). \citep{2014ApJ...790...54C} provided the photometric redshift catalog we used. The photometric redshifts were determined via SED fitting using up to 17 photometric bands (more details of the SED fitting data and techniques can be found in \citep{2014ApJ...790...54C} and \citep{2016ApJ...825...72A}). 
    
    The photometry of \textit{Herschel}/PACS 100 and 160 $\mu$m detected cluster members was presented by \citet{2016ApJ...825...72A}. We used 160 $\mu$m observations with an approximate sensitivity of $0.5 - 2$ mJy RMS to determine the correlation parameter values. The \textit{Spitzer}/MIPS $24 \mu$m observations have reported $3\sigma$ depths of 156 $\mu$Jy at $z = 1$ and 36 $\mu$Jy at $z = 1.5$ (\citep{2013ApJ...779..138B}).
    We refer to \citet{2016ApJ...825...72A} for a detailed description of the ancillary MIR-FIR photometry, \textit{Spitzer}/IRAC, and \textit{Spitzer}/MIPS imaging. The IRAC and PACS counterparts of spectroscopic redshift cluster members were matched using a search radius of 1$\arcsec$.

    \begin{table*}[h!]
  \centering  
  \caption{Cluster sample with deep \textit{Herschel}/PACS imaging and the VLA radio follow-up with their coordinates, spectroscopic redshift, number of detected sources, and measured RMS values for cluster detection.  }
\begin{tabular}{@{}ccccccc@{}}
\hline
\toprule
Cluster ID        & Short ID & R.A.        & DEC.       & Spectroscopic & Detected Galaxy  &   Measured  \\
                  &          & (J2000)     & (J2000)    & Redshift      &  Count &   RMS ($\mu$Jy) \\ \midrule
ISCS J1432.4+3332 & ID1      & 14:32:29.18 & 33:32:36.0 & 1.113         &  13   & 9.9 \\
ISCS J1434.5+3427 & ID2      & 14:34:30.44 & 34:27:12.3 & 1.238         &  15   & 9.1 \\
ISCS J1429.3+3437 & ID3      & 14:29:18.51 & 34:37:25.8 & 1.262         &  11   & 9.2 \\
ISCS J1432.6+3436 & ID4      & 14:32:38.38 & 34:36:49.0 & 1.350         &  10   & 9.1 \\
ISCS J1434.7+3519 & ID5      & 14:34:46.33 & 35:19:33.5 & 1.374         &  11   & 5.2 \\
ISCS J1432.3+3253 & ID6      & 14:32:18.31 & 32:53:07.8 & 1.396         &  14   & 4.1 \\
ISCS J1425.3+3250 & ID7      & 14:25:18.50 & 32:50:40.5 & 1.400         &  8    & 5.9 \\
ISCS J1438.1+3414 & ID8      & 14:38:08.71 & 34:14:19.2 & 1.413         &  16   & 8.8 \\
ISCS J1431.1+3459 & ID9      & 14:31:08.06 & 34:59:43.3 & 1.463         &  7    & 5.4 \\
ISCS J1432.4+3250 & ID10     & 14:32:24.16 & 32:50:03.7 & 1.487         &  17   & 4.1\\
IDCS J1426.5+3508 & ID11     & 14:26:32.95 & 35:08:23.6 & 1.75          &   7   & 4.8 \\ \bottomrule
\end{tabular}
\label{targets_list}
\end{table*}

    The cluster membership was identified by \citet{2008ApJ...684..905E}. Galaxies with spectroscopic redshifts that were considered cluster members needed to have a spectroscopic redshift within 2 Mpc of the cluster center and within 2000 km/s of the systemic cluster velocity. The membership of galaxies with photometric redshifts was determined via constraints on the integral of the normalized probability distribution function. A more detailed description can be found in \citet{2016ApJ...825...72A}.
    
    \subsection{New VLA data reduction and source extraction}

   The radio images for each cluster were obtained with the VLA, centered on the coordinates of the individual galaxy clusters in Table \ref{targets_list}, with the exception of clusters 6 and 10, which were observed in a single image due to their close proximity to one another on the sky. All ten images featuring the 11 clusters were obtained in the 3 GHz (10 cm) band in the B configuration with the S band (2-4 GHz). The estimated RMS sensitivities range from 2-7 $\mu$Jy per beam. The clusters were observed for a total of 23.09 hours of awarded time during October 2017 - May 2018. We performed the data reduction of all images using the Common Astronomy Software Applications (CASA; \citep{2007ASPC..376..127M}) according to the following steps: 1) calibration and flagging of data using the VLA data reduction pipeline version 5.4.2-5 (\citep{2022PASP..134k4501C}); 2) identification and removal of any portions of the data that were corrupted by strong radio frequency interference;  and 3) creation of a science-quality radio-frequency image with the task TCLEAN.

    We performed the task TCLEAN on each image individually initially with the default TCLEAN parameters. We then adjusted the parameters as necessary to obtain the image with the flattest possible background. Suitable values of the cell size parameter range from one-fifth to one-third of the synthesized beam obtained from a publicly available table of configuration properties\footnote{\footnotesize https://science.nrao.edu} from the National Radio Astronomy Observatory (NRAO). For clusters ID1, ID5, and ID8, we therefore set the cell size parameter to 0.42 arcseconds. The primary beam size was equal to $ \frac{\lambda}{D} \times 3600 \arcsec$, where $\lambda$ is an observed wavelength in meters, and D is the VLA diameter in meters. In this case, the primary beam size of all of our images is approximately 18 arcmin. The image size such as the length or width in pixels can be calculated roughly from the primary beam size divided by the cell size. Our final science-image size was therefore  approximately 3000 $\times$ 3000 pixels. We performed a Briggs weighting using a multiterm (multiscale) multifrequency synthesis (MT-MFS) deconvolver with 30000 iterations and a default robustness parameter of 0.5. ID7 was analyzed in a similar manner as for ID1, ID5, and ID8, but we used a robustness parameter of -0.2 to improve the uniformity of the background flux. 

    ID2 was first reduced to produce a science image using the same parameters as for ID1, but the data reduction did not produce as high a quality of the science image because an additional background structure is associated with the bright source within the image. We therefore applied larger cell sizes of 0.55 and 0.7, which are still within the standard range to produce science images. We found that an image with a cell size of 0.55 arcseconds and a lower robust parameter of -0.1 produced the most uniform background.

    For ID3, a large artifact from an extremely bright source lies at the bottom of the image and could not easily be removed. After many attempts to produce a high-quality science image with a uniform background, we were still unable to smooth out the fringe patterns associated with the bright source. Therefore, a nontraditional calibration method of phase-only self-calibration was applied. The image needs to have a signal-to-noise ratio (S/N) higher than 100 at peak. The standard parameters were used with a robustness parameter of 0. 
    
    For ID4, ID6, and ID10, we experimented with the standard parameter values. We found that cell sizes equal to 0.7 arcseconds gave the best result (less background structure), together with a robustness parameter equal to -0.2. Clusters ID6 and ID10 are located in the same image. We experimented with the parameters for the science-image production for ID9 in the same manner as for ID4, ID6, and ID10. However, the best robustness parameter for ID9 was 0.1. This robustness value produces the smoothest possible background and the easiest identification of radio sources.

    ID11 is the most difficult image to calibrate as there is a very bright source right at the center of the image. We therefore performed a self-calibration of each of the four data sets and then combined them into a single science image. The choice to perform a self-calibration on ID11 images was driven by the fact that the S/N of each schedule block was high enough for a phase-only self-calibration to likely be adequate for a science-quality image production. We found that self-calibration generally improved the quality of the resulting science image. The standard parameters were used with a cell size of 0.7 arcseconds. 
    
    All final 3 GHz images had an RMS noise at the pointing center of approximately $4-10$ $\mu$Jy per beam. We decomposed radio sources from the maps using the Python blob detector and source finder (PyBDSF; \citep{2015ascl.soft02007M}). PyBDSF is a software tool designed to decompose images into a set of Gaussians, shapelets, or wavelets and then individual sources. Gaussian fitting was performed simultaneously on islands of sources together with estimations of their uncertainties, using criteria designed to identify and distinguish among separate sources within the island. We performed PyBDSF on all images using a source-detection threshold (the threshold for the island peak in the number of sigma over the mean, $thresh_{pix}$) parameter greater than 0.5 at the IR prior positions. 

    \subsection{Sensitivity comparison}

    Since the estimated depths of the radio images are in the range of 2.0-7.0 $\mu$Jy, we found that our observed data did not reach the expected RMS values. We therefore performed a 3 GHz flux density prediction using SED templates of SFGs to calculate the upper limits of the flux density. We first scaled the observed flux density of IR-detected sources into their rest frame wavelength. Then, we read out the predicted radio 3 GHz flux density at the IR rest frame wavelength using the template from \citet{2009ApJ...692..556R}. Our sample of 129 cluster member galaxies includes 78 detections in radio 3GHz. We used the template to calculate upper limits in radio flux for the remaining 51 radio-undetected galaxies.

\subsection{GOODS-S sample data}

    The Great Observatories Origins Deep Survey-South (GOODS-S) provided deep observations including 24 and 160 $\mu$m. \citep{2020ApJ...901..168A} provided a catalog of the AGN identification in
the GOODS-S/\textit{Hubble} Ultra Deep Field (HUDF) and deep 3 GHz radio data. The 24 and 160 $\mu$m GOODS-S images have a 5$\sigma$ depth of 20 $\mu$Jy and 3$\sigma$ of 2.4 mJy, respectively (\citep{2015A&A...573A..45M}; \citep{2011A&A...533A.119E}). The 3 GHz image of GOODS-S has an RMS noise at the pointing center of 0.75 $\mu$Jy beam$^{-1}$. Since the GOODS-S depth is equivalent to that of our data, we also selected data from the same redshift range as our clusters in order to produce a field galaxy comparison sample in which the clusters matched in redshift and image depth.

\subsection{Catalog demographics}

\begin{figure}[h]
  \resizebox{\hsize}{!}{\includegraphics{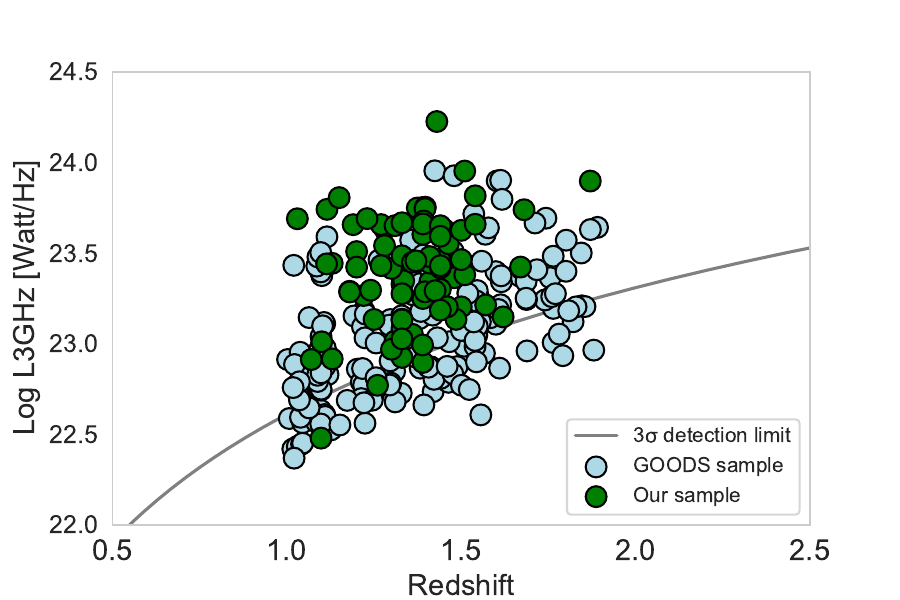}}
  \caption{Rest-frame 3 GHz luminosities of our sample as a function of redshift shown as filled green circles. The luminosities of the GOODS sample are represented as filled blue circles. The solid gray line indicates the 3$\sigma$ detection limit.}
  \label{fig:lum_vs_z}
\end{figure}

    The 3 GHz luminosities are shown in Fig. \ref{fig:lum_vs_z} as a function of redshift for our 129 sources. Our sample contains sources with fluxes in the range of $L_{3GHZ} \sim 10^{22.5} - 10^{24.5}$ WHz$^{-1}$ that lie in a redshift range of 1-2. One hundred and eleven sources are detected at 24 and 160 $\mu$m, respectively. One hundred and ninety GOODS-S data are also shown in Fig. \ref{fig:lum_vs_z}.

\section{AGN identification}
\label{AGN_iden}

    Our main objective is to understand how the environment affects the star formation of galaxies that result in different ratios of the emission in the IR and radio, that is, different values of the correlation slope parameter q. The presence of AGNs in cluster galaxies can contribute substantially to the emission of electromagnetic radiation from the galaxies, especially at short wavelengths. There is also evidence that AGNs coevolve with their host galaxies (\citep{2013arXiv1308.6483K}; \citep{2023A&A...672A..98M}) and are linked through one or more physical processes, such as AGN feedback and other forms of secular evolution in galaxies. Therefore, it is crucial to identify AGNs and exclude them from the SFGs to correct for any effects they may have on the average cluster galaxy q-values. The results from \citet{2022ApJ...941..191L} suggest that there is no one specific method that could identify all types of AGNs. Therefore, several selection methods were applied here. 
    
    We performed an IRAC 4-band selection as described by \citet{2012ApJ...748..142D}. The results from the selection are shown in Col. 6 (DonleyFlag). We also performed an X-ray luminosity selection using X-ray data from \citep{2020ApJS..251....2M}. The results are presented in the last column (New X-ray). We used a rest-frame X-ray luminosity cutoff value of $10^{42.5}$ ergs$^{-1}$. We adopted the AGN selection results from \citep{2016ApJ...825...72A}, who quantified the ratio of the UV-MIR luminosity from the host galaxy ($L_{gal}$) and the total UV-MIR luminosity ($L_{total}$) from the combination of the host galaxy and the AGN via SED fitting in order to identify the host galaxy and AGN contributions shown in Cols. 2-5 (Fgal). This ratio is defined as $F_{gal}$ ($L_{gal}/L_{total}$), where $F_{gal} = 0.5$ is the division into AGN and SFGs. A galaxy with a greater luminosity contribution from the AGN than from its host galaxy has $F_{gal} < 0.5$. The FIR selection was adopted from \citet{2013ApJ...763..123K} and is presented in Col. 4 (K13FIR). The previously identified X-ray AGNs from \citet{2016ApJ...825...72A} are also presented in the second to last column (X-ray) of the table.

 \begin{table*}[h!]
  \centering  
  \caption{Sources that satisfy one or more criterion for AGN identification.}
\begin{tabular}{@{}ccccccccc@{}}
\hline
\toprule
Galaxy ID & Cluster ID & z & K13FIR & Fgal & DonleyFlag & X-ray &  New X-ray \\
    \midrule

J143228.9+333040  & 1  & 1.111  & x  &    &    &    &   &  \\
J143235.8+333631  & 1  & 1.114  & x  &    &    &    &   &  \\
J143232.3+333533  & 1  & 1.115  & x  &    &    &    &   &  \\

J143430.6+342757  & 2  & 1.24  & x  & x  & x  & x  & x  &  \\

J143433.9+352047  & 5  & 1.37  &   &   &   &   & x  &  \\
J143445.7+351921  & 5  & 1.374  & x  &    &    &    &   &  \\
J143501.7+351844  & 5  & 1.39  &   &   &   &   & x  &  \\
J143450.7+351924  & 5  & 1.44  &   &   &   &   & x  &  \\

J143214.6+325551  & 6  & 1.33  &   &   &   &   & x  &  \\
J143216.5+325433  & 6  & 1.392  & x  & x  & x  & x  & x  &  \\
J143218.1+325322  & 6  & 1.396  & x  &    &    &    &   &  \\
J143218.1+325315  & 6  & 1.400  & x  &    &    &    &   &  \\
J143216.5+325224  & 6  & 1.41  & x  & x  &    &    &   &  \\

J142514.1+324940  & 7  & 1.396  & x  &    &    &    &   &  \\

J143808.1+341453  & 8  & 1.400  & x  &    &    & x  &   &  \\
J143806.9+341424  & 8  & 1.425  & x  & x  &    &    &   &  \\
J143817.1+341627  & 8  & 1.43  &   &   &   &   & x  &  \\
J143823.5+341221  & 8  & 1.44  &   &   &   &   & x  &  \\
J143800.6+341712  & 8  & 1.51  &   &   &   &   & x  &  \\

J143117.9+350215  & 9  & 1.5  &   &   &   &   & x  &  \\

J143210.4+324738  & 10  & 1.47  &   &  & x  &   & x  &  \\
J143212.3+325241  & 10  & 1.51  &   &   &   &   & x  &  \\
J143224.5+325305  & 10  & 1.54  &   &   &   &   & x  &  \\

J142649.1+350948  & 11  & 1.84  &   &   &   &   & x  &  \\

     \bottomrule
\end{tabular}

\label{agn_list}
\end{table*}

\section{IR-radio correlation}
\label{IR_correlation}

    We considered the IR-radio correlation using the
    IR flux densities at 24 $\mu$m and 160 $\mu$m to avoid the model assumptions needed to use the total infrared luminosity. To facilitate the comparison, we followed the definitions of the 24 $\mu$m and 160 $\mu$m to radio correlation parameters by \citet{2020ApJ...901..168A}. We define $q_{24}$ in Eq. \ref{eq_q24}, using the 24 $\mu$m IR flux density and the 1.4 GHz radio flux density,

\begin{equation}
    q_{24} \equiv log(S_{24\mu m}/S_{1.4GHz}).
    \label{eq_q24}
\end{equation}

    However, we note that the $24 \mu$m IR flux density could include a significant AGN contribution that we must account for in our analysis. For this reason, we also used the 160 $\mu$m IR flux density, which has a much smaller AGN contribution, to calculate $ q_{160}$, as shown in Eq. \ref{eq_q160},

\begin{equation}
    q_{160} \equiv log(S_{160\mu m}/S_{6GHz})   \label{eq_q160}.
\end{equation}

    From our 3 GHz radio data, we needed to scale our radio data into 1.4 GHz and 6 GHz flux estimates using the same assumption as \citep{2021A&A...647A.123D}, assuming the relation $S_{\nu} \propto {\nu}^{\alpha}$ , with a spectral index value $\alpha = -0.75$. The assumption was discussed and tested by \citet{2021A&A...647A.123D} by comparing scaled 1.4 GHz data (obtained from 3 GHz data) and ancillary 1.4 GHz data. The result suggested that it is reasonable to use $\alpha = -0.75$ as an assumption for the spectral index covering the full stellar mass range.   \\

\begin{figure*}[h!]
\begin{minipage}[t]{0.5\textwidth}
  \includegraphics[width=\linewidth]{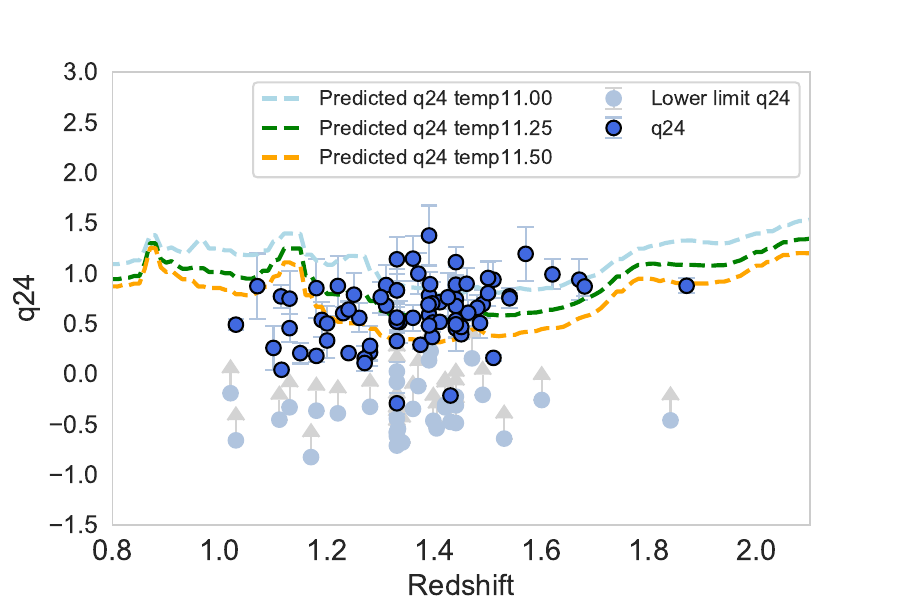}
\end{minipage}
\hfill 
\begin{minipage}[t]{0.5\textwidth}
  \includegraphics[width=\linewidth]{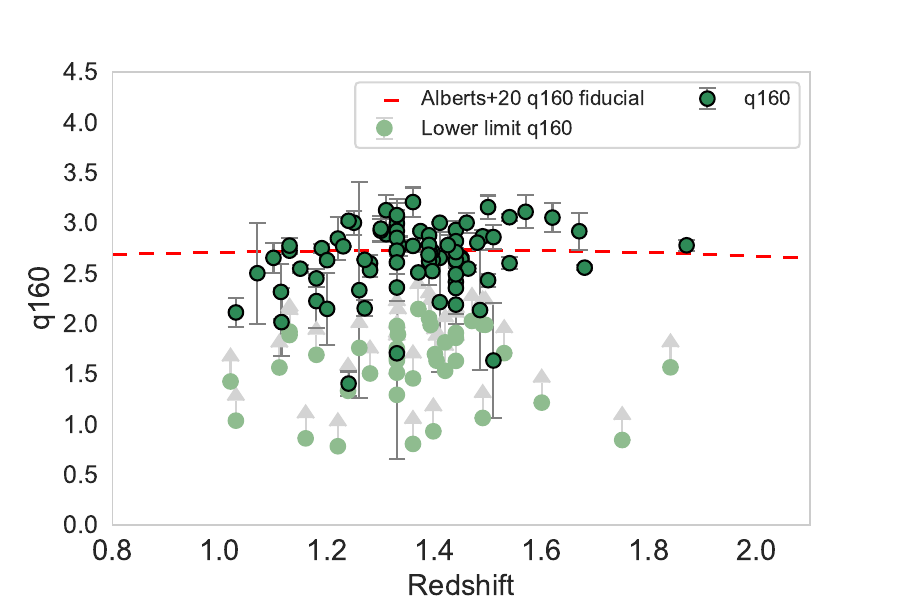}
\end{minipage}
\caption{IR-radio correlation values of the full sample of 11 massive galaxy clusters vs. redshift. Sources with detections in the IR and radio were used to determine the values of q$_{24}$ (left) and q$_{160}$ (right) that are shown as blue and green points, respectively. Sources that were only detected in the IR are lower limits and are shown as light blue and light green points with upper gray arrows. The error bars were derived from the propagation errors. The dashed sky blue, green, and orange lines indicate representative SFG templates from \citep{2009ApJ...692..556R} at $log (L_{IR}/L_\odot\ ) = [11.00, 11.25, 11.50]$ for q$_{24}$. The dashed red line represents the fiducial q$_{160}$ from \citep{2020ApJ...901..168A}.}
  \label{fig:qvalue}
\end{figure*}

    Figure \ref{fig:qvalue} shows q$_{24}$ and q$_{160}$  of all 129 galaxies from 11 clusters in our sample as a function of redshift. Our sample mostly contains data at redshifts z $\sim 1 - 1.6$ since we did not obtain that many detections in the $z = 1.75$ cluster. Since the radio observation did not reach the expected sensitivity, we have 78 secure detections in 3 GHz radio. The 3$\sigma$ lower limits on q$_{24}$ and q$_{160}$ are presented in the plots. We estimated the propagation of the flux measurement uncertainty on the q$_{24}$ and q$_{160}$ data points. The prediction lines of q$_{24}$ and q$_{160}$ are also shown in the figures.

\subsection{Comparison with the GOODS-S sample}

\label{compare_goods}

\begin{figure*}[h!]
\begin{minipage}[t]{0.5\textwidth}
  \includegraphics[width=\linewidth]{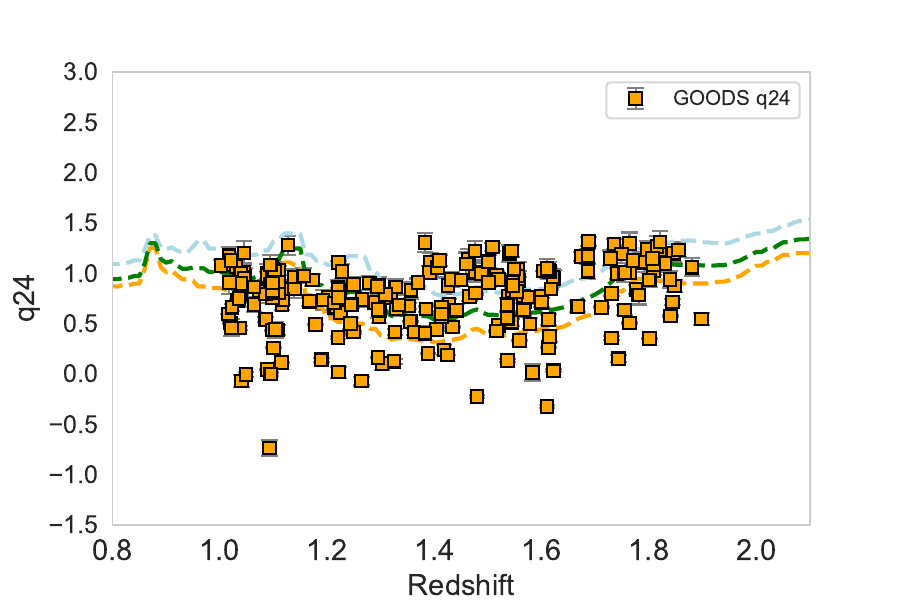}
\end{minipage}
\hfill 
\begin{minipage}[t]{0.5\textwidth}
  \includegraphics[width=\linewidth]{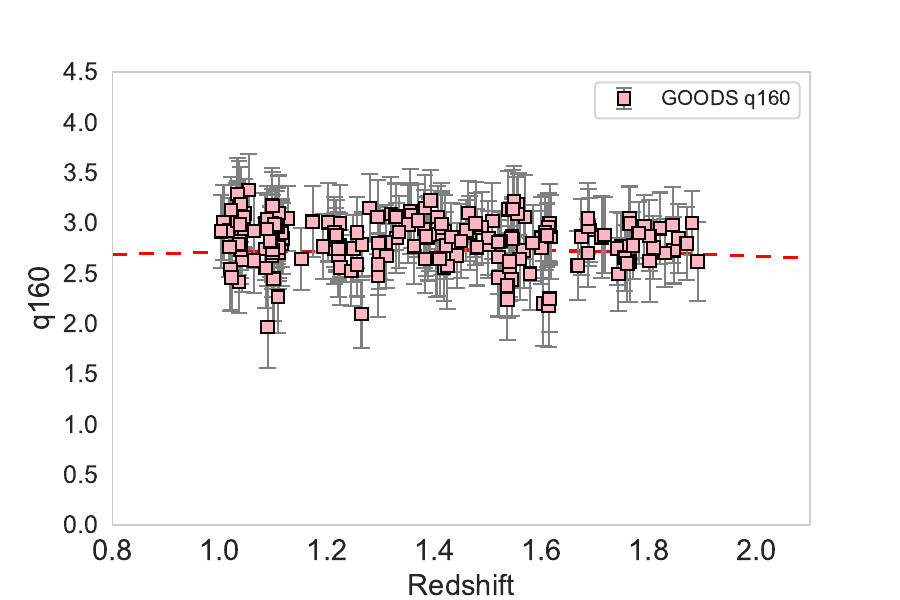}
\end{minipage}
\caption{q$_{24}$ and q$_{160}$ of the GOODS-S field galaxy comparison sample are represented by filled orange and pink squares, respectively. The flux ratios were calculated using the same method as applied to our sample. }
  \label{fig:q_twosamples}
\end{figure*}

In order to understand whether the environment affects the IR-radio correlation, we compared our cluster galaxies to a sample of field galaxies. We selected a comparable field sample with an equivalent depth and the same redshift range as our cluster data from GOODS-S. Equivalent data from \textit{Herschel} PACS 160 $\mu$m and VLA 3 GHz were used to compute the q-values (IR-radio correlation parameters) as shown in Fig. \ref{fig:q_twosamples}. For the q$_{24}$ plot, we excluded 24 AGNs from our sample and 40 AGNs from the GOODS-S sample. In our sample, we identified AGNs using the identification methods described in Sect. \ref{AGN_iden}, and the AGN catalogs for the GOODS-S were drawn from \citep{2022ApJ...941..191L}. We did not exclude AGNs in the q$_{160}$ plot for either of the samples since emission at 160 $\mu$m contributes less to the AGN than 24 $\mu$m (\citep{2017FrASS...4...35P}). 

To compare the two samples statistically, we performed two-sample Kolmogorov–Smirnov (KS) tests. The KS test returns a p value that corresponds to a confidence level in rejecting the null hypothesis of both samples being drawn from the same parent subsample. For example, a common p-value threshold to define significant differences between the two samples compared in the test is 0.05, which corresponds to a 95\% confidence.

The KS test applied to the q$_{24}$ distributions for the cluster and field galaxies returned a p value of 0.02, which marginally rejects the null hypothesis that both samples are drawn from identical parent samples with a 98\% confidence ($\sim 2.3 
 \ \sigma$). This indicates the need for a larger sample to determine whether there are genuine differences in the ratio of the 24 $\mu$m to radio between cluster and field galaxies.

\begin{figure*}[ht!]
\begin{minipage}[t]{0.33\textwidth}
  \includegraphics[width=\linewidth]{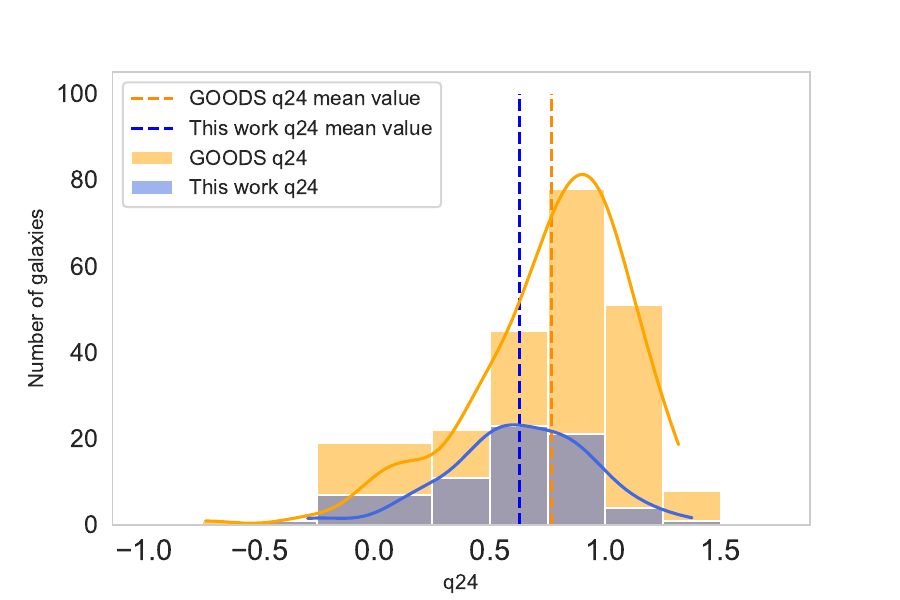}
  
\end{minipage}
\hfill 
\begin{minipage}[t]{0.33\textwidth}
  \includegraphics[width=\linewidth]{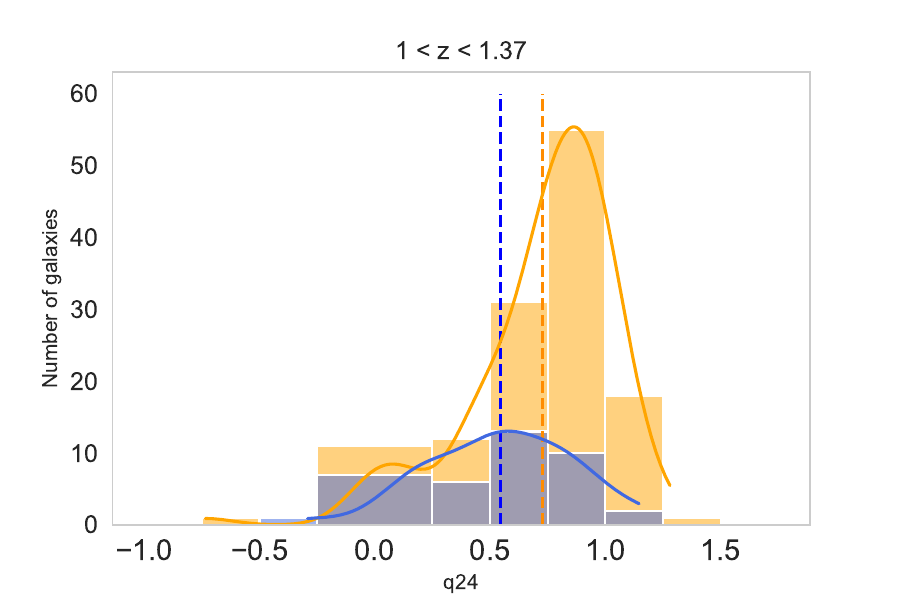}
  
  \label{fig:second}
\end{minipage}
\hfill
\begin{minipage}[t]{0.33\textwidth}
  \includegraphics[width=\linewidth]{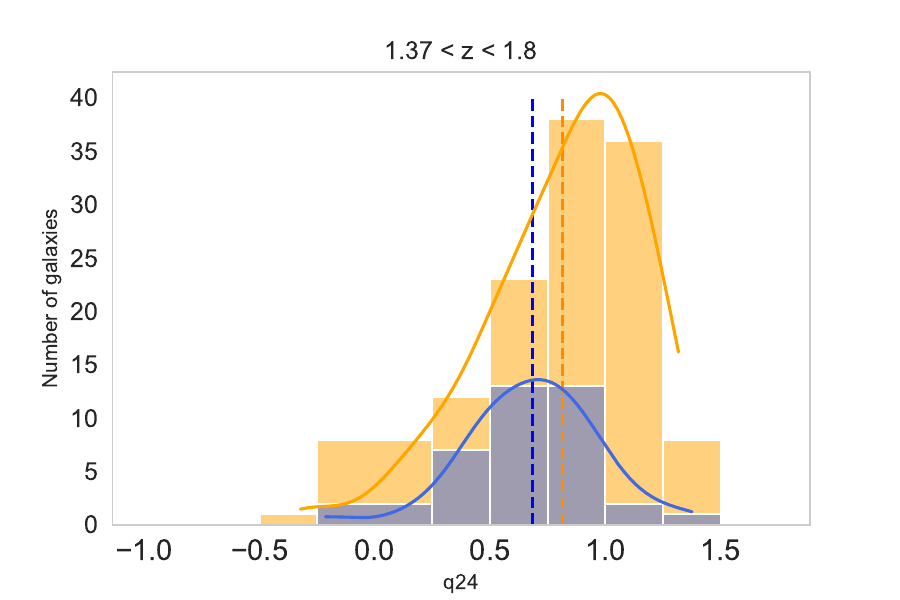}
\end{minipage}%

\caption{Histograms of the q$_{24}$ values of the entire selected sample (left), the low-redshift sample (middle), and the high-redshift sample (right). The dashed blue lines represent the mean values of q$_{24}$ that belong to our subsamples, and the dashed orange lines show the mean values of q$_{24}$ from the GOODS-S subsamples.}
\label{fig:q24_hist}
\end{figure*}

We plot the histograms of the q$_{24}$ values in Fig. \ref{fig:q24_hist} for the whole selected sample and in redshift bins. Our low-redshift bin corresponds to galaxies with redshifts from 1 to 1.37, and our high-redshift bin corresponds to redshifts 1.37 to 1.95. The redshift separation value 1.37 was adopted from \citep{2016ApJ...825...72A}. The mean values of q$_{24}$ from this work are lower than the q$_{24}$ mean values from GOODS-S in the entire selected sample and in the two subsamples. To confirm this, we also performed a two-sample KS test of two redshift-based subsamples, as shown in the middle and right panel of Fig. \ref{fig:q24_hist}. 

The results of the KS tests in the low- and high-redshift bins are 0.03 and 0.07, respectively. The low-redshift p value of 0.03 corresponds to a 97\% confidence level, which is $\sim 2.2 \ \sigma$. The high-redshift p value of 0.07 corresponds to a 93\% confidence level \textbf{($\sim 1.8 \ \sigma$)}. As with the full sample, in the low-redshift sample, we can marginally reject the null hypothesis of identical parent distributions of cluster and field q$_{24}$ values with a 97\% confidence ($\sim 2.2 \ \sigma$). A larger sample might allow us to distinguish more clearly between the cases of identical or distinct parent distributions for q$_{24}$ values at low redshift.  On the other hand, our KS test results are consistent with the null hypothesis in the case of higher redshift, with a confidence of only 93\% ($\sim 1.8 \ \sigma$) in the likelihood of distinct parent distributions. In both cases, the confidence in the similarities and differences in the q$_{24}$ distributions is limited by the sample size, which in the redshift subsamples is smaller than in the full sample.

\begin{figure*}[ht!]
\begin{minipage}[t]{0.33\textwidth}
  \includegraphics[width=\linewidth]{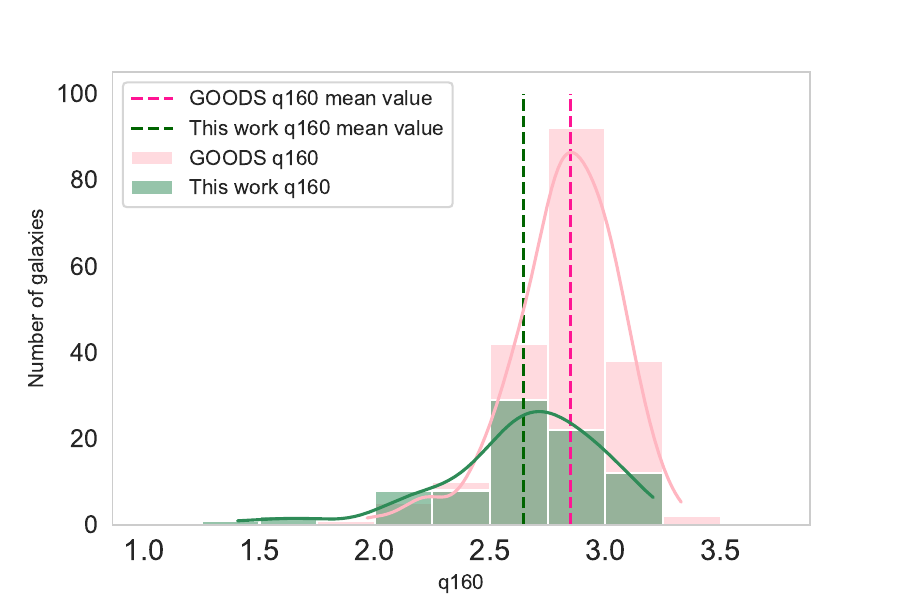}
  
\end{minipage}
\hfill 
\begin{minipage}[t]{0.33\textwidth}
  \includegraphics[width=\linewidth]{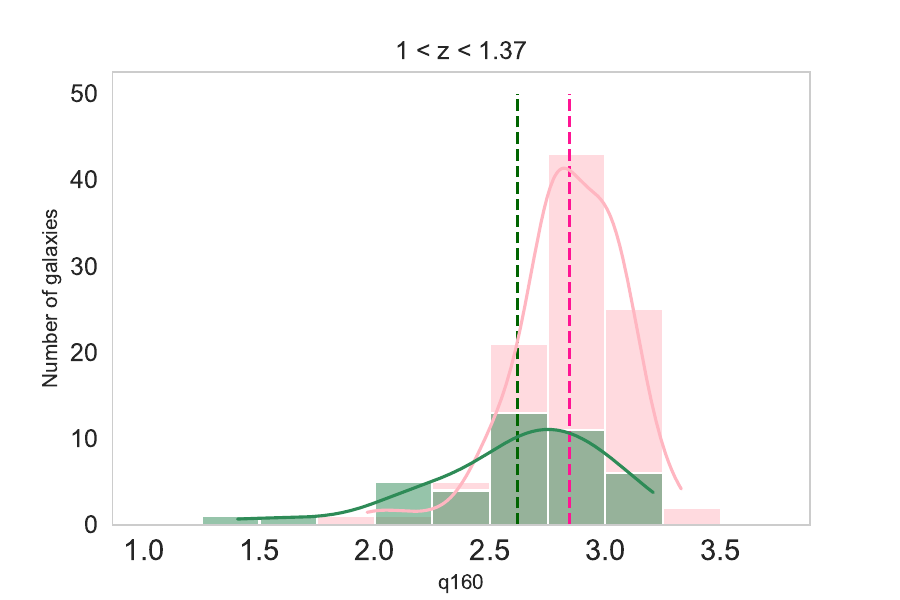}
  
  \label{fig:second}
\end{minipage}
\hfill
\begin{minipage}[t]{0.33\textwidth}
  \includegraphics[width=\linewidth]{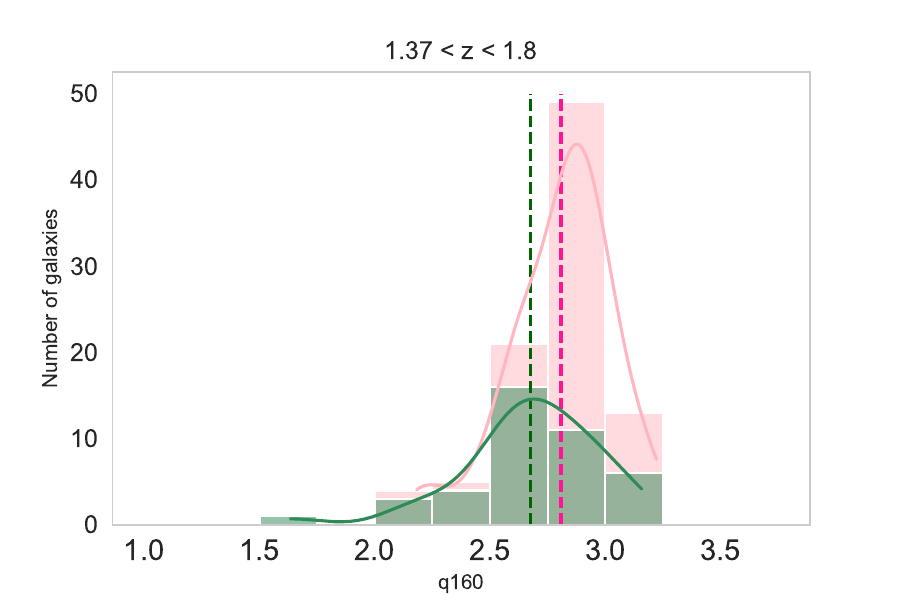}
\end{minipage}%

\caption{Histograms of the q$_{160}$  values of the entire selected sample (left), the low-redshift sample (middle), and the high-redshift sample (right). The dashed green lines represent the mean values of q$_{160}$ in our cluster galaxy subsamples. The dashed pink lines show the mean values of q$_{160}$ from the GOODS-S subsamples.}
\label{fig:q160_hist}
\end{figure*}

For q$_{160}$, the KS test gives a p value of 0.01, which we consider to be a marginal rejection with a 99\% confidence ($\sim 2.5 \ \sigma$). This suggests a likely significant difference in the parent distributions of the q$_{160}$ values for cluster and field galaxies, but larger samples are needed to be more certain of this result.

As performed above with q$_{24}$, we divided q$_{160}$ into two subsamples by redshift, and we show the histograms in Fig. \ref{fig:q160_hist}. Using the corresponding null hypothesis for the two-sample KS test, we found that the p values are 0.04 and 0.10 for the low- and high-redshift bin, respectively. Consistent with q$_{24}$, we obtain a p value of 0.04, which corresponds to a 96\% confidence level, which is $\sim 2 \ \sigma$. We consider this a marginal rejection with a confidence of 96\%. This indicates a notable probability of distinct parent distributions for cluster and field q$_{160}$, but our sample size is inadequate for high confidence. For the high-redshift subsample, we obtain a p value of 0.10, which is a 90\% confidence level or $\sim 1.6 \ \sigma$. This result is consistent with the null hypothesis of identical parent distributions of q$_{160}$ at high redshift, although our sample size limits our confidence, as with all other subsamples and our full samples.

\subsection{q values as a function of projected radius}

\begin{figure*}[ht!]
\begin{minipage}[t]{0.5\textwidth}
  \includegraphics[width=\linewidth]{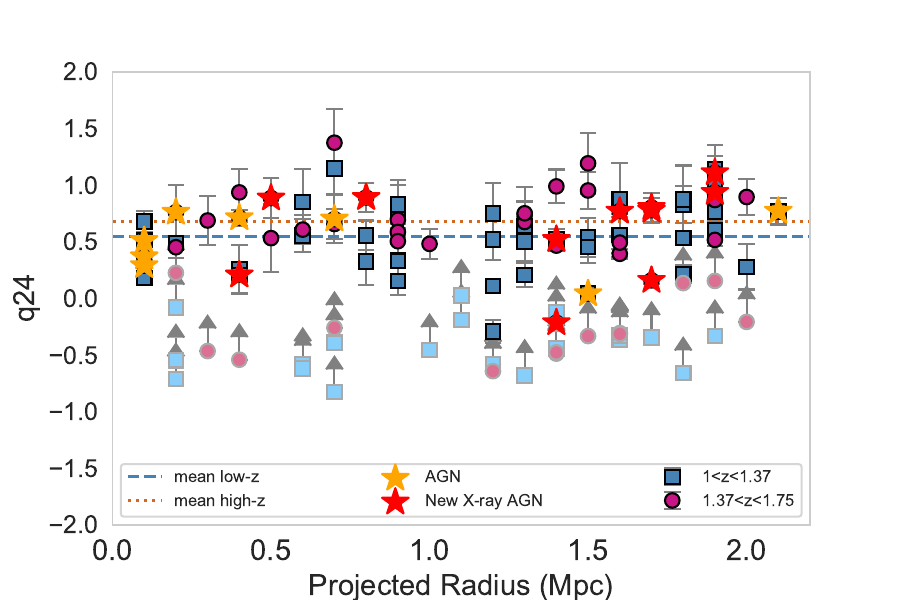}
\end{minipage}
\hfill 
\begin{minipage}[t]{0.5\textwidth}
  \includegraphics[width=\linewidth]{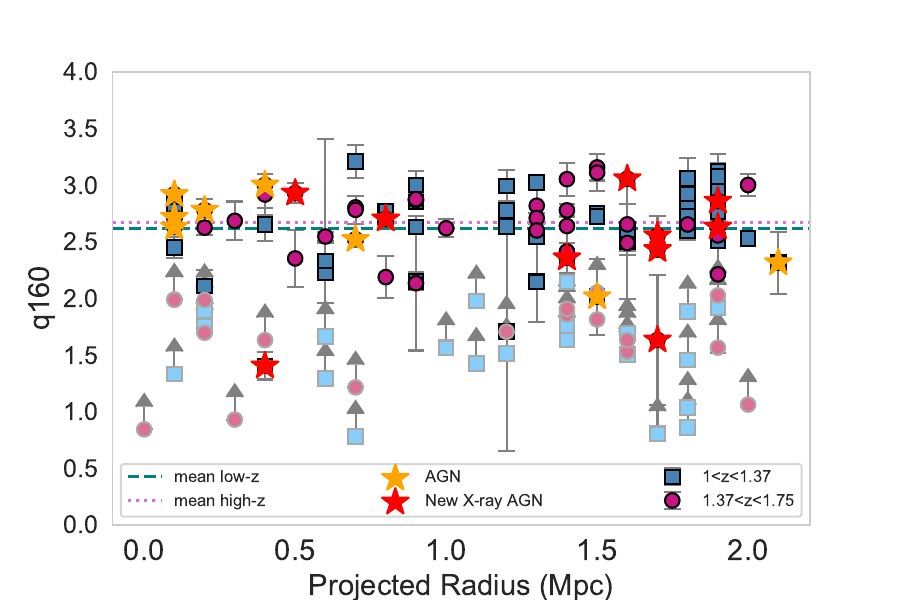}
\end{minipage}
\caption{q$_{24}$  (left) and q$_{160}$  (right) values of two redshift bins as a function of R. Low-z galaxies are plotted as filled blue squares, and high-z galaxies are plotted as filled violet circles. The lower limit of the q values of the low-z sample is plotted as filled light blue squares, and the lower limit of the high-z sample is plotted as filled light violet circles. Previously identified AGNs are overplotted with yellow stars, and newly identified AGNs are plotted with red stars. The mean correlation parameter values of the low- and high-redshift subsamples are shown with dashed and dotted lines, respectively. }
  \label{fig:q_ccr}
\end{figure*}

In addition to comparing the global IR-radio correlation between galaxies in clusters and field, we considered the variations in the local density by plotting in Fig. \ref{fig:q_ccr} the IR-radio the IR-radio correlation for the cluster sample as a function of the galaxy projected radius, R (in Mpc). We divided our sample into two groups: galaxies with R<1Mpc, the inner subsample; and galaxies with R>1Mpc, the outer subsample. We chose the cutoff value of 1Mpc for the projected radius since it is approximately the virial radius for clusters in this mass range. The virial radius is typically assumed to be the boundary of the cluster. Galaxies outside this radius are in the outskirts and are thus considered less likely to have been exposed to the intense environmental processes of the galaxy cluster for a significant amount of time. Testing the null hypothesis that there is no difference in q values based on these two subsamples, we find that the KS tests give p values of 0.28 and 0.35 for q$_{24}$  and q$_{160}$ , respectively. This suggests that we found evidence for identical parent distributions of the correlation parameter values as as a function of local environment between the inner and outer galaxies of both q$_{24}$  and q$_{160}$  with a 72\% confidence ($\sim 1 \ \sigma$) and  65\% confidence ($\sim 1 \ \sigma$), respectively. Although our sample size is limited and the exact virial radii of the clusters were not calculated, these low p values are highly suggestive of the null hypothesis of identical parent distributions.

\subsection{q values as a function of redshift}

Although our complete sample contains a relatively small number of galaxies but it spans 1.6 Gyr in cosmic time. We tested if the q values that plotted as a function of projected radius from cluster center are different in two sub-samples. With the null hypothesis being that there is no difference in the parent distribution of the q values based on different samples, we discovered that the KS tests give 0.14 and 0.32 p values for q$_{24}$ and q$_{160}$, respectively. This means that we found no evidence of a variation in the q-parameters as a function of projected cluster-centric radius between low and high redshift for q$_{24}$ and q$_{160}$. We note that this may not be definitive evidence of identical parent distributions because the samples we used for this analysis are among the smallest we produced for comparison with our dataset.

\section{Presence of AGNs}

\label{presence_AGN}

After finding only marginal evidence of environmental effects on the IR-radio correlation when comparing cluster to field galaxies and as a function of projected cluster-centric radius, we investigated whether the presence of AGNs affects the star-forming galaxy q values enough to reduce the significance of the results of the K-S tests of q$_{24}$. We took 12 previously identified AGNs and 12 newly identified AGNs into account using new Chandra data as explained in Sect. \ref{AGN_iden}. A visual inspection suggests that the AGN population is indistinguishable from the non-AGN population. A KS test was performed
between AGN hosts and normal SFGs. With the null hypothesis being that there is no difference in q values based on these two sub-samples, the KS tests give 0.46 and 0.53 p values for q$_{24}$ and q$_{160}$ , respectively. This suggests that we found no evidence for an effect on the average q values due to the presence of AGNs. We note that all of our AGNs are radio quiet, which may cause their 24 $\mu$m IR and radio emission to be dominated by star formation (\citep{2020ApJ...901..168A}). 

\begin{figure*}[ht!]
\begin{minipage}[t]{0.5\textwidth}
  \includegraphics[width=\linewidth]{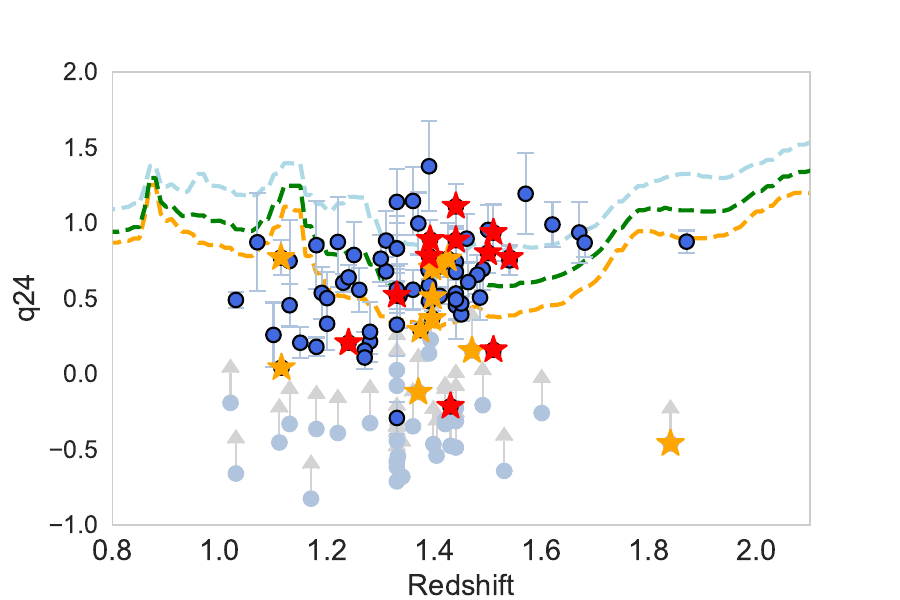}
\end{minipage}
\hfill 
\begin{minipage}[t]{0.5\textwidth}
  \includegraphics[width=\linewidth]{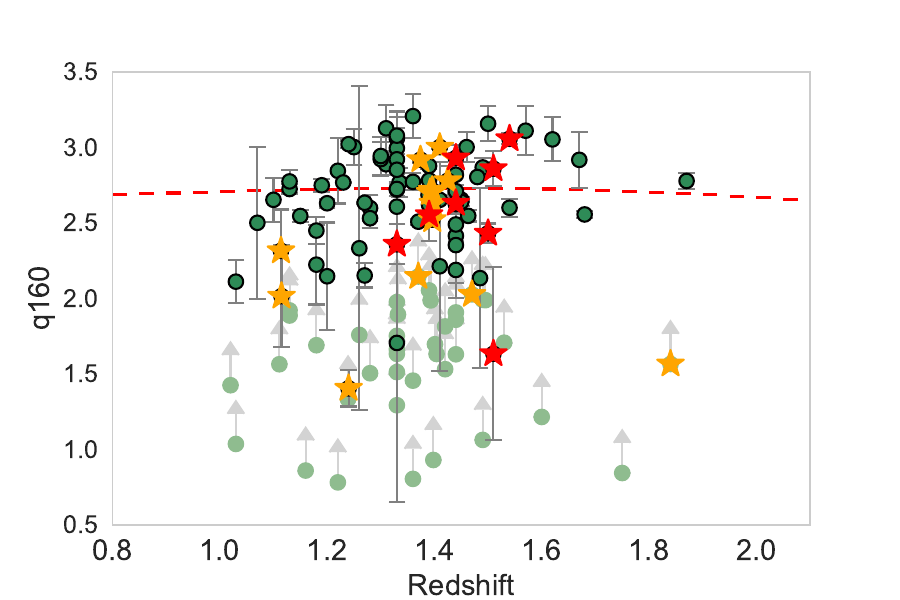}
\end{minipage}
\caption{IR-radio correlation values of the full sample of 11 massive galaxy clusters vs. redshift for q$_{24}$ (left) and q$_{160}$ (right). The symbols and lines are the same as in Fig. \ref{fig:qvalue}. Previously identified AGNs are overplotted with yellow stars, and newly identified AGNs are plotted with red stars.  }
  \label{fig:q_agn}
\end{figure*}

To understand the distribution of AGNs in redshift space, we overplot them using stars in Fig. \ref{fig:q_ccr} and Fig. \ref{fig:q_agn}. A large portion of our AGNs are concentrated at z $\sim 1.4$. Thirty percent of the galaxies in the z$\sim$1.4 clusters are AGN (6 are newly identified in this work, and only 50\% lie in lower-redshift clusters). This is consistent with studies that showed an increasing fraction of AGN in clusters at high redshift.

\section{Discussion}
\label{discussion}

The IR-radio correlation has often been studied since the 1970s. One of the first studies involving galaxy clusters was \citet{2001AJ....121.1903M}. The authors performed an extensive analysis of cluster environmental effects on the IR-radio correlation in local Abell galaxy clusters. They found that the correlation is tight for SFGs from the centers of clusters to their classical Abell radii. They found an excess of radio emission in SFGs that was small, but statistically significant. They found no significant correlation between the IR-radio correlation and the line-of-sight velocity relative to the cluster systemic velocities. They rejected the hypothesis that AGNs contribute to an increase in the radio emission and supported the alternative hypothesis that the cluster environment affects the IR-radio correlation values. 

The results reported by \citet{2015MNRAS.447..168R}, on the other hand, found no evidence for variability in the IR-radio correlation with a change in redshift or environment. The radio emission excess in high-redshift galaxies in clusters was found to be more common than in lower-redshift galaxies. The explanation was that in the earlier epoch, radio-excess sources are enhanced by the cluster. The authors revealed a consistency between the correlation parameter value of blue galaxies with the predicted value from the field galaxies. Along with the above-mentioned results from analyses of intermediate-redshift clusters, the correlation parameter values of a higher-redshift (z $\sim$ 2) galaxy cluster was studied by \citet{2021MNRAS.503.1174K}. They indicated that the infrared–radio correlation is not affected by being in a dense environment, especially at the core of a cluster. Nevertheless, the studies were based on one single cluster, and further research was therefore needed with more clusters at high redshifts, such as this work, in order to reach clearer conclusions.

We provided the first analysis of the IR-radio correlation of a sample of 11 high-intermediate redshift galaxy clusters with relatively homogeneous data and analytical methods. In Sect. \ref{IR_correlation} we reported the p values from the KS tests of the whole sample for q$_{24}$ and q$_{160}$. We reject the null hypothesis of identical parent distributions of q values for cluster and field galaxies at the 95\% and 99\% significance level, respectively. It is also notable that the mean values of q$_{24}$ and q$_{160}$ of our sample from galaxies that lie in clusters are significantly lower than for GOODS-S field galaxies. These results are consistent with the previously mentioned studies in the local Universe (\citep{1995AJ....109.1582A}, \citep{2004ApJ...600..695R}, \citep{2009ApJ...706..482M}). We also showed that the correlation is tight at this redshift, even in clusters, as there are no substantial outliers from the correlation. 

As we investigated further by dividing our sample into low- and high-redshift bins, we found that the low-redshift subsample showed marginally significant evidence of environmental effects in the results of the KS tests in both q$_{24}$  and q$_{160}$. As demonstrated in Fig. \ref{fig:q24_hist} and \ref{fig:q160_hist}, the mean values of q$_{24}$ and q$_{160}$ in the low-redshift bin are lower in cluster samples than in the field galaxy sample from GOODS-S, indicating radio excess in cluster galaxies due to effects of the dense environment. The high-redshift subsample showed results that are consistent with the null hypothesis, that is, an insignificant probability of distinct parent distributions, for both q$_{24}$ and q$_{160}$. This suggests that the environment likely does not affect the correlation values at high redshift, consistent with the single high-redshift cluster result by \citet{2021MNRAS.503.1174K} mentioned above. 

We found that evidence of environmental effects in the lower-redshift sample is inconsistent with the result from the intermediate-redshift study of \citet{2015MNRAS.447..168R}. Therefore, we showed for the first time evidence of environmental effects on the IR-radio correlation at z $\sim 1-1.4$, which is consistent with previous evidence of rapid increases in environmental effects on galaxies, which are most commonly seen in the form of environmental quenching, after z $\sim 1.5$ (\citep{2017MNRAS.465L.104N}; \citep{2020MNRAS.499.3061N}). At the high-redshift end of our sample, the lack of an environmental contribution to galaxy star formation and the resulting IR and radio emission is consistent with the results from \citet{2021MNRAS.503.1174K} and is also consistent with the weaker evidence for environmental quenching of star formation in clusters at z $\gtrsim 1.4$. Nonetheless, our understanding of the IR-radio correlation in particular would benefit from further research with larger samples of cluster galaxies at late cosmic noon.

In Sect. \ref{presence_AGN} we showed a tight correlation in the q$_{24}$ and q$_{160}$ values at all distances from the cluster center because we found that the presence of an AGN in our sample has no effect on the mean values of q$_{24}$ and q$_{160}$. This is also consistent with the result reported by \citet{2001AJ....121.1903M}. We noted that all of our identified AGNs are radio quiet. 

Our results report that the IR-radio correlation values of cluster galaxies are lower than field galaxies, indicating the environmental effect. This finding supports the radio excess scenario, which is associated with physical processes that occur in galaxy clusters due to the galaxy evolution, such as mergers and ram pressure stripping. Even though the origin of the radio excess might lie in other processes, such as the presence of AGN, this is ruled out because we did not identify any radio-loud AGN in our sample. Larger samples of galaxy clusters may further aid in clarifying the (un)importance of AGNs for the radio excess phenomenon. 

We lack information regarding mergers of galaxies in our sample, but it is doubtful that the effect of mergers is strongly significant in this case because our results show systematically lower q values. Therefore, further observation, specifically at higher resolution, and investigation is strongly encouraged. Ram pressure stripping, on the other hand, might cause the radio excess. Further investigation is needed to confirm this. 

Another possible physical cause for excess radio emission resulting in lower q values in clusters than in the field is cluster-scale diffuse radio emission in galaxy clusters. \citet{2024A&A...686A..82B} recently showed evidence for a link between galaxy IR and radio emission and cluster-scale diffuse radio emission in a galaxy cluster. Emission associated with radio excess can also be found in radio relics, which are produced by shocks when galaxy clusters merge (\citep{2024A&A...686A..55L}). 

In addition to observational work conducted to investigate the FIR-radio correlation, several models were proposed to explain the result. Generally, our results cannot be explained by any existing model, as there is a lack of factor(s) concerning all the environmental effects. 
\citet{2025ApJ...980..135P} presented simulations of the IR-radio correlation that also considered the effect of AGN feedback. The simulation verified the predictions of the calorimeter theories proposed by \citet{1989A&A...218...67V}. FIR emission is assumed to originate from energetic (ionizing) photons, while nonthermal radio continuum emission is assumed to arise from relativistic electrons. Both emissions are assumed to be proportional to the SNe rate, assuming that the energy sources for the FIR emission and the radiating dust grains are in instantaneous communication. We find that our results might be explained by the simulations proposed by \citet{2025ApJ...980..135P} if other environmental effects, such as ram pressure stripping or mergers are considered. We note that our definition of the q parameter is different from the mentioned work, and their conclusions on the effect of the AGN feedback are also different. 

Regarding the IR-radio correlation values as a function of projected radius to understand variations in the local density, we find that our result implies an acceptance of the null hypothesis of identical parent distributions, meaning that two subsamples are not different. This might suggest that the redshift effect is very subtle for this particular redshift range and sample. 

We finally plotted the q values as a function of redshift that spans 1.6 Gyr. We found no difference in the parent distribution of q values based on different samples, which means that we are not able to confirm the redshift effect from our sample. We suggest that further investigation with a larger dataset is strongly needed.

\section{Conclusions}

We investigated the IR-radio correlation in massive galaxy clusters using existing IR (24 and 160 $\mu$m) data and new VLA 3 GHz data. The KS test was used to compare the IR-Radio correlation
values, q$_{24}$ and q$_{160}$, for galaxies in our cluster sample and in the GOODS-S sample at an equivalent depth. AGNs were identified in the cluster as well as in the control sample in order to explore their impact on the IR-radio correlation. Our main conclusions are listed below.

\begin{enumerate}

      \item We observe a difference at medium to high confidence level ($\sim2\ \sigma-3\  \sigma$) between the IR-radio correlation parameter values of our full galaxy cluster sample and the galaxy field control sample. 
      
      \item Based on splitting the full galaxy cluster sample into low-redshift (1<z<1.37) and high-redshift (1.37<z<1.8) subsamples, we conclude that the difference observed in the IR-Radio correlation values with respect to the field galaxy control sample is driven by the low-redshift subsample. In the low-redshift subsample, the mean q values of the galaxy cluster sample are significantly lower than the mean q values from the galaxy field control sample. We interpret this result as being related either causally or by a shared common cause to the cluster star formation quenching.
      
      \item We found that in higher-redshift clusters, there is no significant difference in the IR-radio correlation values between cluster galaxies and the field galaxy control sample. This agrees with the weakening in environmental effects (e.g., quenching) at $z \gtrsim 1.4$.
      
      \item Our results indicate no evidence of any difference in the IR-radio correlation between galaxies closer to the cluster center (R < 1 Mpc) and galaxies in the cluster outskirts (R > 1 Mpc).
      
      \item We found no difference in the IR-radio correlation values in our comparison between normal SFGs and radio-quiet AGNs. This suggests that the presence of radio-quiet AGN does not affect the correlation parameter values. 

      \item Our results suggest that the radio excess scenario that is associated with physical processes that occur in galaxy clusters due to the galaxy evolution such as mergers and ram pressure stripping causes the lower correlation values in galaxy clusters.

   \end{enumerate}

\begin{acknowledgements}
      The National Radio Astronomy Observatory is a facility of the National Science Foundation operated under cooperative agreement by Associated Universities, Inc. This work is based on observations made with \textit{Herschel}, a European Space Agency Cornerstone Mission with significant participation by NASA.  This work is additionally based on observations made with \textit{Spitzer}, which was operated by the Jet Propulsion Laboratory, California Institute of Technology under contract with NASA. Part of this work was supported by the German
      \emph{Deut\-sche For\-schungs\-ge\-mein\-schaft, DFG\/} project
      number Ts~17/2--1. SA acknowledges support from the JWST Mid-Infrared Instrument (MIRI) Science Team Lead, grant 80NSSC18K0555, from NASA Goddard Space Flight Center to the University of Arizona. This work was also funded by the National Agency for Research and Development (ANID) / Scholarship Program / DOCTORADO BECAS CHILE / 2022 - 21211581. 
\end{acknowledgements}

\bibliographystyle{aa}
\bibliography{aa52996-24.bib}

%
%

\end{document}